\documentstyle[amssym]{mn}

\font\japit = cmti10 at 10truept
\input{epsf.sty}

\def\eps@scaling{1}				
\def\epsscale#1{\def\eps@scaling{#1}}
\def\epsfsize#1#2{\eps@scaling#1}

\title
     [Decorrelating the Power Spectrum]
{\vglue-3.0truecm
\centerline{\japit Published in Monthly Notices}
\vglue 2.5truecm
\noindent
Decorrelating the Power Spectrum of Galaxies
\author[A. J. S. Hamilton \& M. Tegmark]
     {A. J. S. Hamilton$^1$ \& Max Tegmark$^2$ \\
	$^1$JILA and Dept.\ Astrophysical \& Planetary Sciences,
	Box 440, U. Colorado, Boulder CO 80309, USA; \\
	\ Andrew.Hamilton@Colorado.EDU; http:$/\!/$casa.colorado.edu/$\sim$ajsh \\
        $^2$Institute for Advanced Study, Princeton, NJ 08540, USA;
        max@ias.edu; http:$/\!/$www.sns.ias.edu/$\sim$max}
}

\newcommand{\rmn}{\rm}
\newcommand{\bmi}{\bmath}

\newcommand{\mx}{\sf}		

\newcommand{\be}{\begin{equation}}
\newcommand{\ee}{\end{equation}}
\newcommand{\ba}{\begin{eqnarray}}
\newcommand{\ea}{\end{eqnarray}}

\newcommand{\dd}{{\rmn d}}	
\newcommand{\e}{{\rmn e}}	

\newcommand{\transpose}{\top}

\newcommand{\n}{{\bmi n}}

\newcommand{\1}{{\mx 1}}
\newcommand{\bB}{{\mx B}}

\newcommand{\bE}{{\mx E}}
\newcommand{\bF}{{\mx F}}
\newcommand{\bG}{{\mx G}}
\newcommand{\bL}{{\mx L}}

\newcommand{\bO}{{\mx O}}

\newcommand{\bU}{{\mx U}}
\newcommand{\bV}{{\mx V}}
\newcommand{\bW}{{\mx W}}
\newcommand{\bX}{{\mx X}}
\newcommand{\bLambda}{{\mx\Lambda}}

\newcommand{\Mpc}{{\rmn Mpc}}
\newcommand{\dex}{{\rmn dex}}
\newcommand{\km}{{\rmn km}}

\newcommand{\sr}{{\rmn sr}}

\newcommand{\PI}{\pi}

\newcommand{\apj}[2]{ApJ, #1, #2}
\newcommand{\apjs}[2]{ApJ Supp, #1, #2}

\newcommand{\mn}[2]{MNRAS, #1, #2}
\newcommand{\pr}[2]{Phys.\ Rev., #1, #2}
\newcommand{\prl}[2]{PRL, #1, #2}

\newcommand{\Ffig}{
    \begin{figure}
    \begin{center}
    \leavevmode
    \epsscale{.8}
    \epsfbox{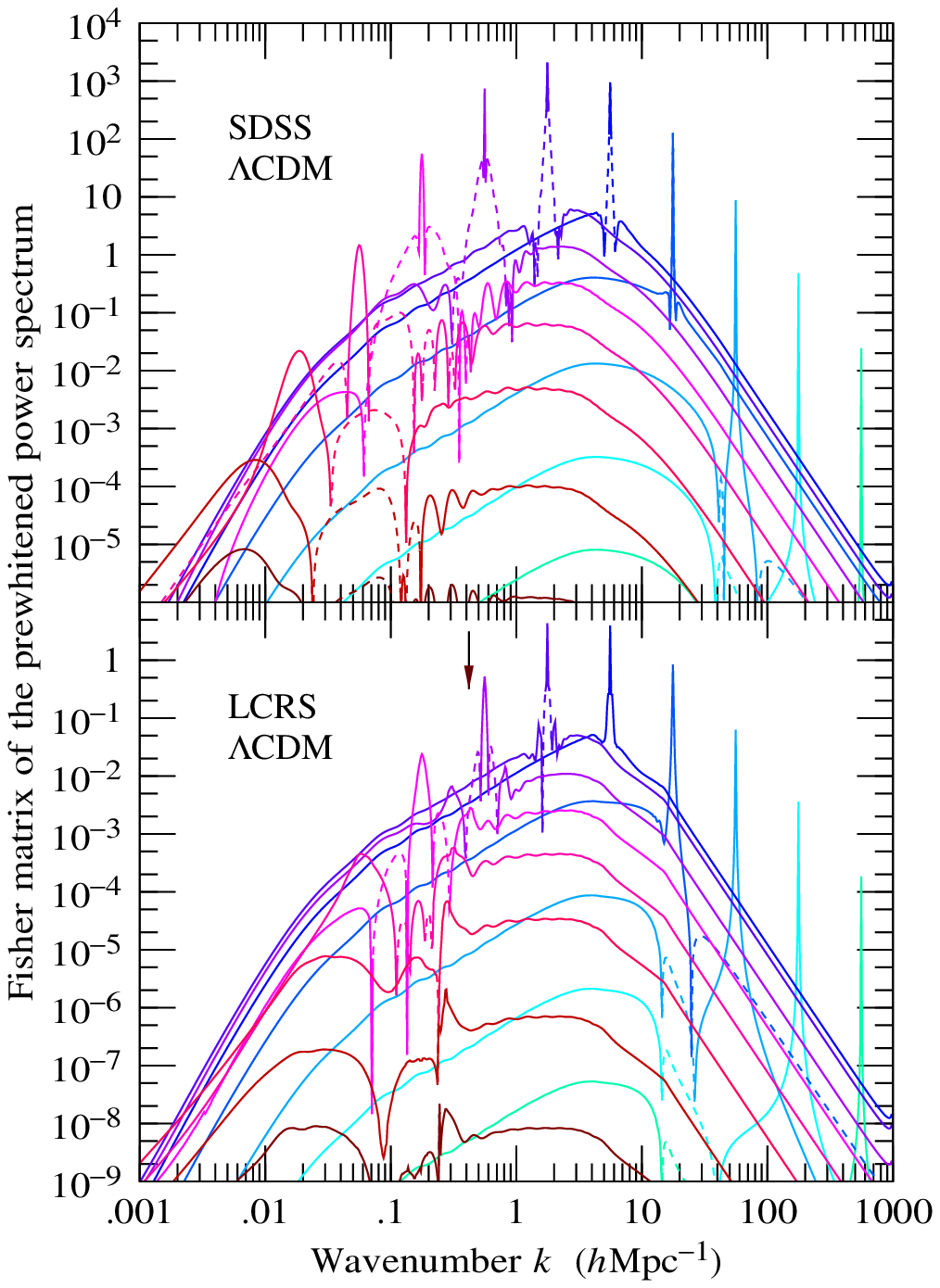}
    \end{center}
    \caption[1]{
Fisher information matrix $\bG$
of the scaled prewhitened power spectra $\hat\bX/\bX$ of
(top) SDSS and (bottom) LCRS.
The Fisher matrix was computed in the FKP approximation,
in the manner described in \S6 of Paper~3.
The vertical arrow in the lower panel indicates the wavenumber
$k = \PI/7.5 \  h \, \Mpc^{-1}$,
where the thickness of a $1\fdg5$ slice at median depth equals a
half-wavelength, below which the FKP approximation may fail in LCRS.
The assumed prior power spectrum is an
observationally concordant $\Lambda$CDM model from
Eisenstein \& Hu (1998),
nonlinearly evolved according to the formula of Peacock \& Dodds (1996).
Each line represents a row (or column, since the matrix is symmetric)
of the Fisher matrix;
each row peaks at, or at small $k$ near, the diagonal.
Lines are dashed where the Fisher matrix is negative.
The resolution is $\Delta\log k = 1/128$.
    \label{F}
    }
    \end{figure}
}

\newcommand{\Xpcdsdssfig}{
    \begin{figure}
    \begin{center}
    \leavevmode
    \epsscale{.8}
    \epsfbox{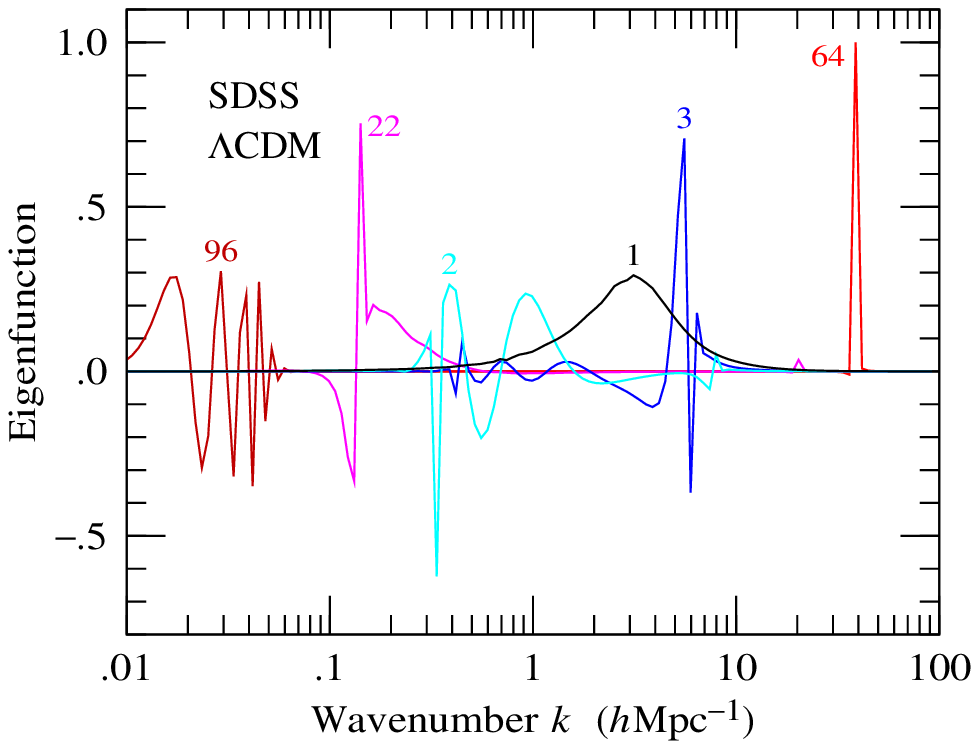}
    \end{center}
    \caption[1]{
Principal component decomposition of the Fisher matrix $\bG$
of the scaled prewhitened power spectrum $\hat\bX/\bX$ of SDSS,
for a $\Lambda$CDM prior power spectrum.
Each band-power window is an orthonormal eigenfunction of $\bG$,
or equivalently an eigenfunction of the covariance
$\bX^{-1} \langle \Delta \hat\bX \, \Delta \hat\bX^\transpose \rangle \bX^{-1}$.
The eigenfunctions are labelled in order of the expected variances
of the corresponding band-power,
the lowest numbered eigenfunction having the smallest variance.
The Fisher matrix here was computed over the range
$k = 0.01$--$100 \, h\, \Mpc^{-1}$,
at a resolution of $\Delta\log k = 1/32$.
    \label{Xpcd_sdss}
    }
    \end{figure}
}

\newcommand{\Xcholfig}{
    \begin{figure}
    \begin{center}
    \leavevmode
    \epsscale{.8}
    \epsfbox{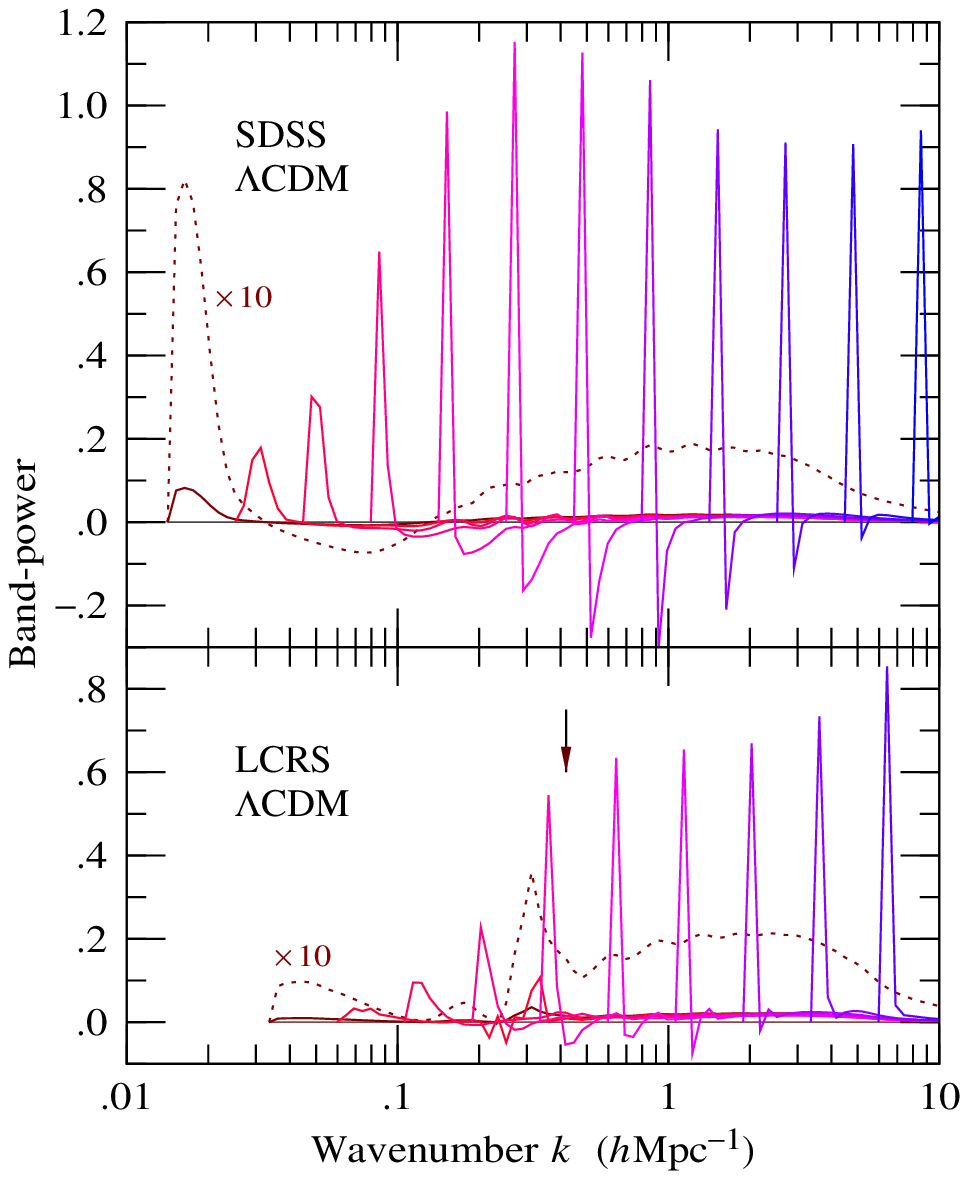}
    \end{center}
    \caption[1]{
Sample of band-power windows from the Cholesky decomposition of
the Fisher matrix $\bG$ of the scaled prewhitened power spectrum
of (top) SDSS and (bottom) LCRS.
The Fisher matrix here was truncated to a maximum wavenumber
of $k_{\max} = 0.015 \, h \, \Mpc^{-1}$ in SDSS,
and $k_{\max} = 0.034 \, h \, \Mpc^{-1}$ in LCRS,
to ensure that all its eigenvalues were positive.
The resolution is $\Delta\log k = 1/32$,
and only every eighth band-power window is plotted.
The band-power window at the smallest wavenumber
is also shown multiplied by a factor of $10$, as a dotted line.
The vertical arrow on the LCRS plot indicates the wavenumber
$k = \PI/7.5 \  h \, \Mpc^{-1}$
below which the FKP approximation may fail in LCRS.
    \label{Xchol}
    }
    \end{figure}
}

\newcommand{\Fhfig}{
    \begin{figure}
    \begin{center}
    \leavevmode
    \epsscale{.8}
    \epsfbox{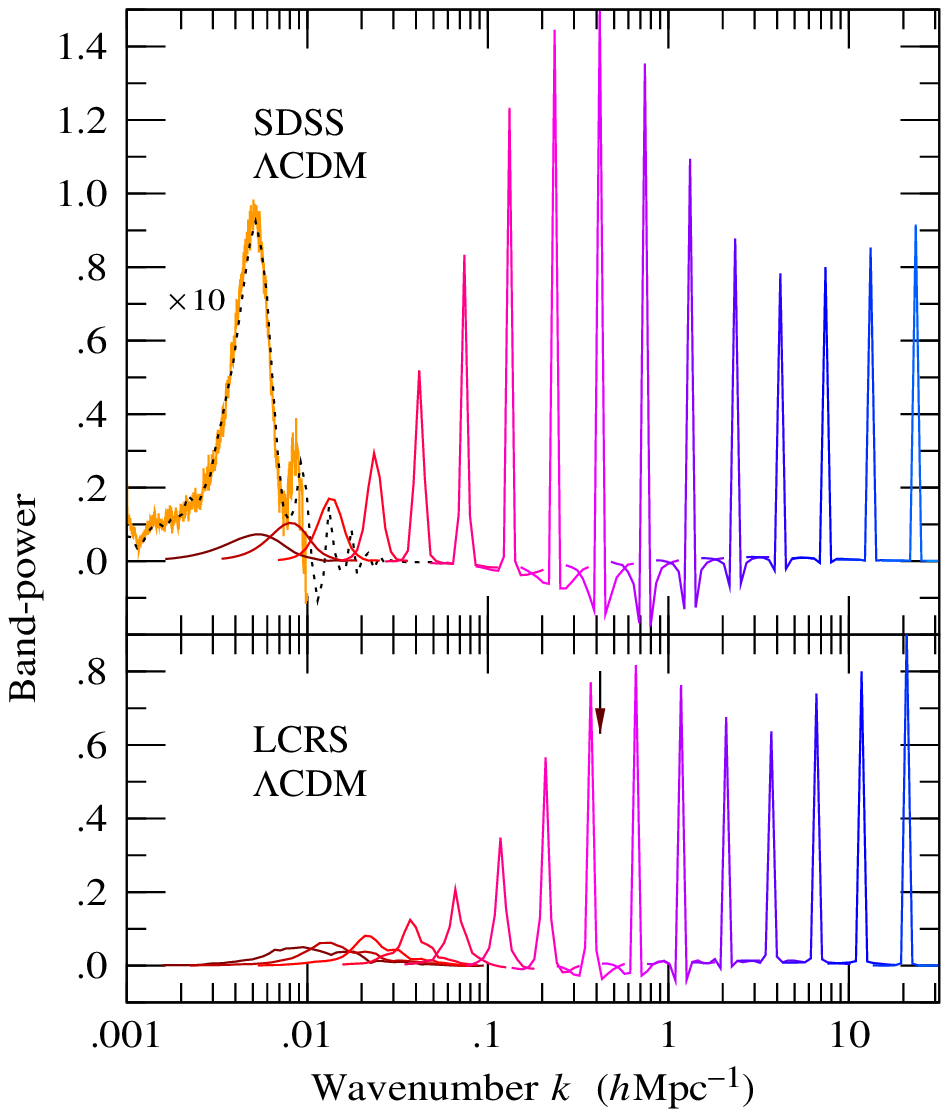}
    \end{center}
    \caption[1]{
Sample of band-power windows from the square root $\bG^{1/2}$ of
the Fisher matrix of the scaled prewhitened power spectrum
of (top) SDSS and (bottom) LCRS.
The range of the matrix is $k = 0.001$--$100 \, h \, \Mpc^{-1}$,
with resolution $\Delta\log k = 1/32$.
Every eighth band-power window is plotted,
starting at $k = 0.0042 \, h \, \Mpc^{-1}$ in SDSS,
and $k = 0.0066 \, h \, \Mpc^{-1}$ in LCRS.
For SDSS, the band-power window at the smallest wavenumber,
a nominal wavenumber of $k = 0.001 \, h \, \Mpc^{-1}$,
is shown multiplied by a factor of $10$, as a dotted line.
What looks like a shaded curve (but is actually just a finely oscillating line)
underlying this last band-power is effectively the same band-power,
at the same nominal wavenumber of $k = 0.001 \, h \, \Mpc^{-1}$,
but computed at $32$ times higher resolution, $\Delta\log k = 1/1024$,
over the range $k = 0.001$--$0.01 \, h \, \Mpc^{-1}$
(the range was restricted to keep the computational size down).
The high-resolution band-power is multiplied by $320$
to raise it to the same level as the low-resolution band-power
(recall that the band-powers here are normalized so that the sum
of their elements is one).
The vertical arrow on the LCRS plot indicates the wavenumber
$k = \PI/7.5 \  h \, \Mpc^{-1}$
below which the FKP approximation may fail in LCRS.
    \label{Fh}
    }
    \end{figure}
}

\newcommand{\Fhlinfig}{
    \begin{figure}
    \begin{center}
    \leavevmode
    \epsscale{.8}
    \epsfbox{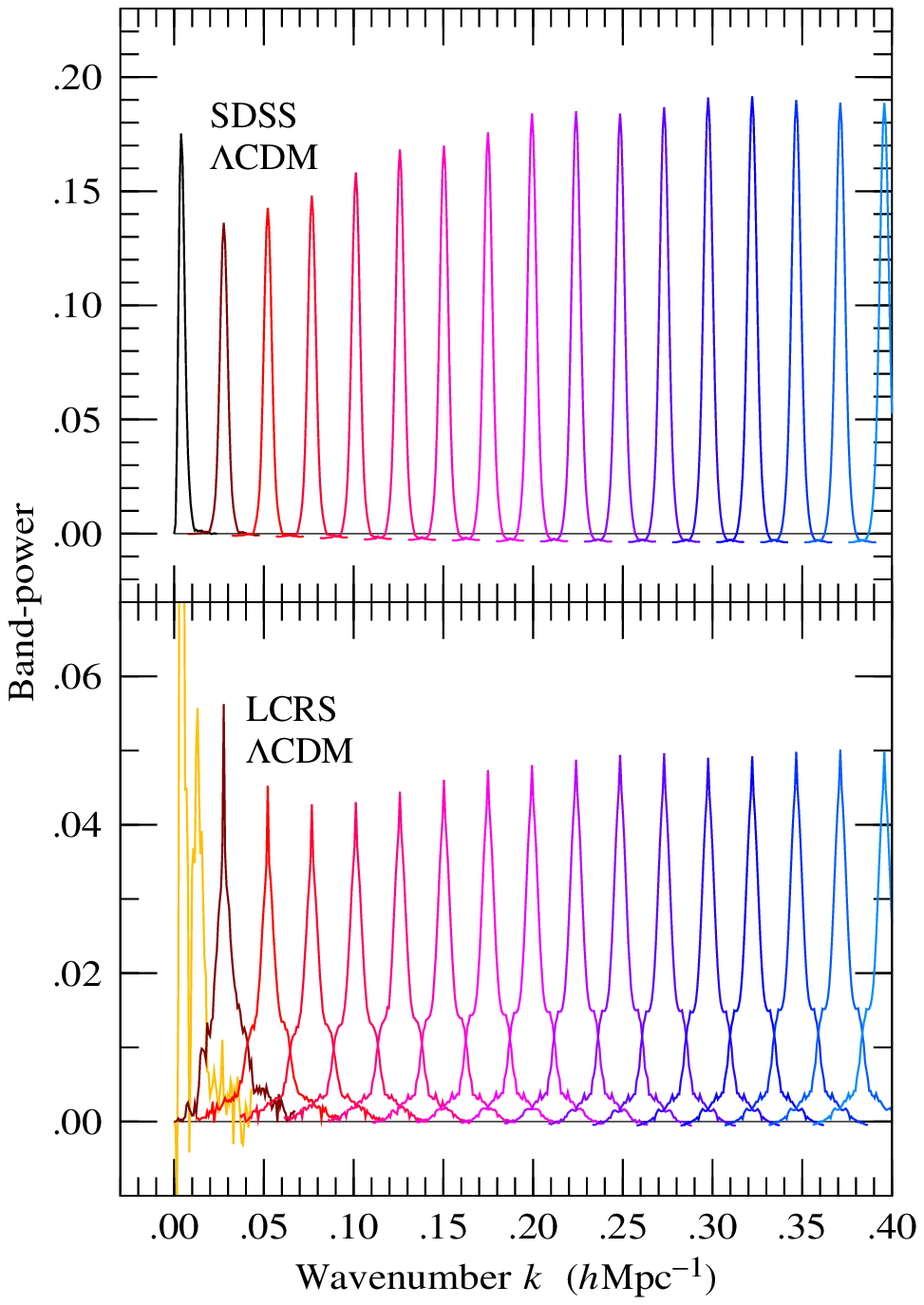}
    \end{center}
    \caption[1]{
Sample of band-power windows from the square root $\bG^{1/2}$ of
the Fisher matrix of the scaled prewhitened power spectrum
of (top) SDSS and (bottom) LCRS,
with linear rather than logarithmic gridding in wavenumber.
The resolution is $\Delta k = \PI/4096 \  h\, \Mpc^{-1}$,
finer than the natural resolution of the survey
by a factor of about 4 in SDSS, 6 in LCRS.
Every 32nd band-power is plotted,
starting at a nominal wavenumber of $k = \PI/1024 \  h\, \Mpc^{-1}$.
For LCRS, the first band-power plotted,
at nominal wavenumber $k = \PI/1024 \  h\, \Mpc^{-1}$,
is jittery.
For LRCS, the FKP approximation may fail at all wavenumbers shown.
    \label{Fh_lin}
    }
    \end{figure}
}

\newcommand{\Fhgfig}{
    \begin{figure}
    \begin{center}
    \leavevmode
    \epsscale{.8}
    \epsfbox{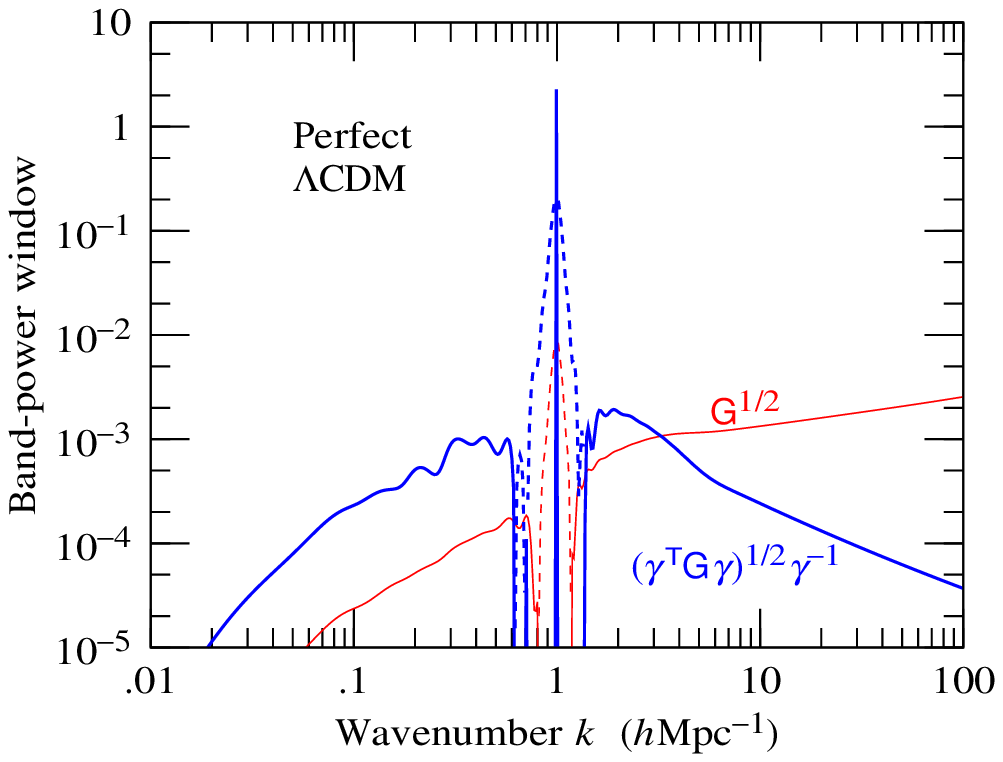}
    \end{center}
    \caption[1]{
Comparison of band-power windows obtained from
(thin line) the square root of the Fisher matrix, $\bG^{1/2}$,
versus (thick line) the scaled square root of the Fisher matrix,
$( \gamma^\transpose \bG \, \gamma )^{1/2} \gamma^{-1}$
with scaling function $\gamma(k) = k^{3/2}$,
for the case of a perfect, noiseless survey.
Only the single band-power centred at $k = 1 \, \Mpc^{-1}$ is shown.
Lines are dashed where the windows are negative.
The resolution is $\Delta\log k = 1/128$.
The figure illustrates that in cases where the Fisher matrix varies steeply,
scaling the Fisher matrix before taking its square root can
help to produce more symmetric band-powers.
    \label{Fhg}
    }
    \end{figure}
}

\newcommand{\xiafig}{
    \begin{figure}
    \begin{center}
    \leavevmode
    \epsscale{.8}
    \epsfbox{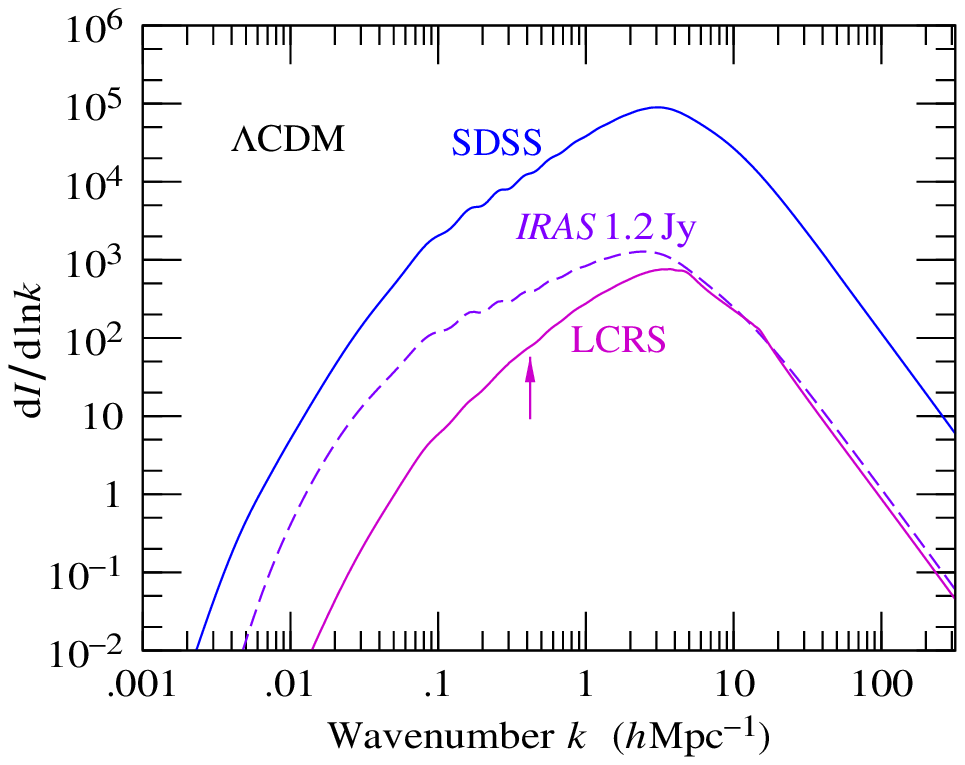}
    \end{center}
    \caption[1]{
Inverse variance, or information,
per logarithmic interval of wavenumber
in the scaled prewhitened power spectra
of the SDSS, LCRS, and {\it IRAS\/}~1.2~Jy surveys,
for the $\Lambda$CDM prior power spectrum.
This information is the inverse variance on each band-power,
if the band-powers are an $\e$-fold wide.
The information in the LCRS is probably underestimated here,
especially at larger scales, because of the assumption of the FKP approximation.
The vertical arrow indicates the wavenumber
$k = \PI/7.5 \  h \, \Mpc^{-1}$
below which the FKP approximation may fail in LCRS.
    \label{xia}
    }
    \end{figure}
}

\newcommand{\xikfig}{
    \begin{figure}
    \begin{center}
    \leavevmode
    \epsscale{.8}
    \epsfbox{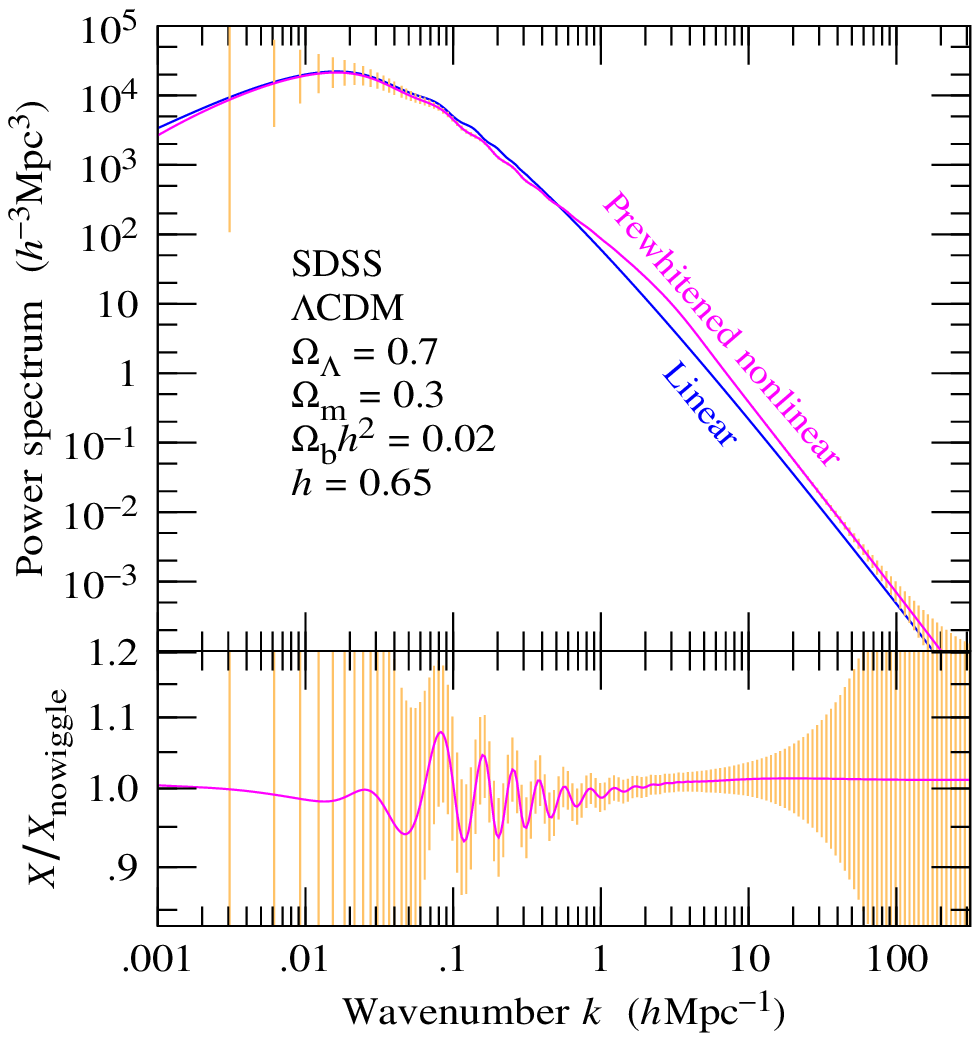}
    \end{center}
    \caption[1]{\small
Decorrelated power spectrum expected from SDSS.
The prior power spectrum is the observationally concordant $\Lambda$CDM model
of Eisenstein \& Hu (1998) used throughout this paper, with
parameters as indicated, nonlinearly evolved according to the formula of
Peacock \& Dodds (1996).
Both the linear power spectrum $\xi_{\rmn L}(k)$ and the prewhitened
nonlinear power spectrum $X(k)$ are shown.
Error bars on the prewhitened nonlinear power spectrum
are linearly spaced in wavenumber at
$\Delta k = \PI/1024 \  h \, \Mpc^{-1}$ up to $k = 0.04 \, h \, \Mpc^{-1}$,
then logarithmically spaced at $\Delta \log k = 1/32$.
Eisenstein \& Hu also provide a `nowiggle' power spectrum that is
a smooth fit through the baryonic wiggles in the spectrum.
The lower panel shows the ratio of the prewhitened nonlinear power spectrum
to the prewhitened nonlinear nowiggle power spectrum.
    \label{xik}
    }
    \end{figure}
}

\begin{document}

\maketitle

\begin{abstract}
It is shown how to decorrelate the (prewhitened) power spectrum measured
from a galaxy survey into a set of high resolution uncorrelated band-powers.
The treatment includes nonlinearity, but not redshift distortions.
Amongst the infinitely many possible decorrelation matrices,
the square root of the Fisher matrix, or a scaled version thereof,
offers a particularly good choice,
in the sense that the band-power windows are narrow,
approximately symmetric,
and well-behaved in the presence of noise. 
We use this method to compute band-power windows for, and the information
content of, the Sloan Digital Sky Survey, the Las Campanas Redshift Survey,
and the IRAS 1.2~Jy Survey.
\end{abstract}



\section{Introduction}
\label{intro}

Accurate measurements of the principal cosmological parameters
now appear to be within reach (Turner 1999).
Large redshift surveys of galaxies,
notably the Two-Degree Field Survey (2dF) (Colless 1998; Folkes et al.\ 1999) 
and the Sloan Digital Sky Survey (SDSS) (Gunn \& Weinberg 1995; Margon 1998), 
should be a gold mine of cosmological information over wavenumbers
$k \sim 0.003$--$300 \, h \, \Mpc^{-1}$.
Information from such surveys 
on large scales ($k \la 0.1\,h\,\Mpc^{-1}$) should
improve greatly the accuracy attainable
from upcoming Cosmic Microwave Background (CMB) experiments alone
(Eisenstein et al.\ 1999),
but in fact most of the information in these surveys is on 
small scales, in the nonlinear regime (Tegmark 1997b).

While the power spectrum may be the ideal carrier
of information at the largest, linear scales
where density fluctuations may well be Gaussian,
it proves less satisfactory at moderate and smaller scales,
since nonlinear evolution induces a broad covariance between
estimates of power at different wavenumbers,
as emphasized by Meiksin \& White (1999)
and Scoccimarro, Zaldarriaga \& Hui (1999).
Ultimately,
one can imagine that there exists some kind of mapping that
translates each independent piece of information in the
(Gaussian) linear power spectrum into some corresponding independent
piece of information in the nonlinear power spectrum,
or some related quantity.
That such a mapping exists at least at some level
is evidenced by the success of the analytic
linear $\leftrightarrow$ nonlinear mapping formulae
of Hamilton et al.\ (1991, hereafter HKLM) and Peacock \& Dodds (1994, 1996).
However, if such a mapping really existed,
that was not only invertible but also,
as in the HKLM-Peacock-Dodds formalism,
mapped delta-functions of linear power at each linear wavenumber
into delta-functions of nonlinear power (or some transformation thereof)
at some possibly different nonlinear wavenumber,
then that mapping ought to translate uncorrelated quantities in the linear
regime -- powers at different wavenumbers --
into uncorrelated quantities in the nonlinear regime.
The fact that the nonlinear power spectrum is broadly correlated
over different wavenumbers shows that the HKLM-Peacock-Dodds
formalism cannot be entirely correct.

In the preceding paper (Hamilton 2000, hereafter Paper~3),
it was shown that prewhitening the nonlinear power spectrum
-- transforming the power spectrum in such a way
that the noise covariance becomes proportional to the unit matrix --
substantially narrows the covariance of power.
Moreover, this narrowing of the covariance of power occurs
for all power spectra tested, including both realistic power spectra
and power law spectra over the full range of indices permitted
by the hierarchical model.
It should be emphasized that these conclusions are premised on
the hierarchical model for the higher order correlations,
with constant hierarchical amplitudes,
and need to be tested with $N$-body simulations.

In the meantime,
if indeed there exists an invertible
linear $\leftrightarrow$ nonlinear mapping of cosmological power spectra,
then it would appear that the prewhitened nonlinear power spectrum
should offer a closer approximation to the right hand side of this mapping
than does the nonlinear power spectrum itself.
Whatever the case,
the prewhitened nonlinear power spectrum has the practical benefit
of enabling the agenda of the present paper
-- decorrelating the galaxy power spectrum --
to succeed over the full range of linear to nonlinear wavenumbers.
That is to say,
if one attempted to decorrelate the nonlinear power spectrum itself into a set
of uncorrelated band-powers, then the band-power windows would be so broad,
with almost canceling positive and negative parts,
that it would be hard to interpret the band-powers as
representing the power spectrum in any meaningful way.
By contrast,
the covariance of the prewhitened nonlinear power is already narrow enough
that decorrelation into band-powers works quite satisfactorily.

Paper~3 showed how to construct
a near approximation to the minimum variance estimator and
Fisher information matrix of the prewhitened nonlinear power spectrum
in the Feldman, Kaiser \& Peacock (1994, hereafter FKP) approximation,
valid at wavelengths short compared to the scale of the survey.
In the present paper we describe how to complete the processing of
the prewhitened nonlinear power spectrum into a set of
decorrelated band-powers that come close to fulfilling the ideals of
(a) being uncorrelated,
(b) having the highest possible resolution,
and (c) having the smallest possible error bars.

The idea of decorrelating the power spectrum was
proposed by Hamilton (1997; hereafter Paper~2),
and was further discussed and successfully applied to the CMB
by Tegmark \& Hamilton (1998).

Like Paper~3, the present paper ignores redshift distortions
and other complicating factors, such as light-to-mass bias,
on the Alfa-Romeo (1969 P159 Shop Manual)
principle that it is best to adjust one thing at a time.

\section{Data}
\label{data}

This paper presents examples mainly for two redshift surveys of galaxies,
the Sloan Digital Sky Survey
(SDSS; Gunn \& Weinberg 1995; Margon 1998)
and the Las Campanas Redshift Survey
(LCRS; Shectman et al.\ 1996;
Lin et al. 1996).
The information content of the prewhitened power spectrum of the
{\it IRAS\/} 1.2~Jy survey
(Fisher et al. 1995)
is also shown for comparison in Figure~\ref{xia}.

Calculating the Fisher matrix of the prewhitened power spectrum,
as described in \S6 of Paper~3,
involves evaluating the FKP-weighted pair integrals $R(r;\mu)$
(Paper~3, eq.~[84]),
commonly denoted $\langle R R \rangle$,
of the survey.
For SDSS and {\it IRAS\/} 1.2~Jy,
the pair integrals $R(r;\mu)$,
were computed in the manner described by Hamilton (1993),
over a logarithmic grid spaced at
$0.02 \, \dex$ in pair separation $r$
and $0.1 \, \dex$ in FKP constant $\mu$.
For LCRS,
we thank Huan Lin (1996, private communication)
for providing FKP-weighted pair integrals $R(r;\mu)$
on a grid of separations $r$ and FKP constants $\mu$.

The geometry of SDSS is described by
Knapp, Lupton \& Strauss (1997, `Survey Strategy' link).
We consider only the main galaxy sample,
not the Bright Red Galaxy sample.
We thank D. H. Weinberg (1997, private communication)
for providing the anticipated radial selection function of the survey,
and M. A. Strauss (1997, private communication)
for a computer file and detailed explanation of the angular mask.
The angular mask of the northern survey consists of 45 overlapping
stripes each $2\fdg5$ wide in declination,
forming a roughly elliptical region extending $110^\circ$ in declination
by $130^\circ$ in the orthogonal direction.
The angular mask of the southern survey consists of 3 separate $2\fdg5$ wide
stripes centred at declinations $15^\circ$, $0^\circ$, and $-10^\circ$.
The survey covers a total area, north plus south, of $3.357 \, \sr$.

The LCRS (Shectman et al.\ 1996)
covers six narrow slices
each approximately $1\fdg5$ in declination by $80^\circ$ in right ascension,
three each in the north and south galactic caps.

The angular mask of the {\it IRAS\/} 1.2~Jy survey
(Fisher et al.\ 1995)
is the same as that of the 2~Jy survey
(Strauss et al.\ 1990, 1992),
and consists of the sky at galactic latitudes $|b| > 5^\circ$,
less 1465 approximately $1^\circ \times 1^\circ$ squares,
an area of $11.0258 \, \sr$ altogether.
We thank M. A. Strauss (1992, private communication)
for a computer listing of the excluded squares.
The radial selection function was computed by the method of Turner (1979).
Distances were computed from redshifts assuming a flat
matter-dominated cosmology,
and galaxy luminosities $L$ were corrected for evolution with redshift $z$
according to $L \propto 1 + z$.

\section{Decorrelation}
\label{decorrelation}

\subsection{Decorrelation matrices}
\label{decorrelationW}

Let $\bF$ be the Fisher information matrix
(see Tegmark, Taylor \& Heavens 1997 for a review)
of a set of estimators $\hat\theta$
of parameters $\theta$ to be measured from observations.
Below we will specialize to the case where the parameters
are the prewhitened power spectrum,
but for the moment the parameters $\theta$ could be anything.
Assume,
thanks to the central limit theorem or otherwise,
that the covariance matrix of the estimators $\hat\theta$ is adequately
approximated by the inverse of the Fisher matrix
\be
  \langle \Delta\hat \theta \Delta\hat \theta^\transpose \rangle
  = \bF^{-1}
  \ .
\ee

A {\it decorrelation matrix} $\bW$ is
any real square matrix, not necessarily orthogonal, satisfying
\be
\label{decorr}
  \bF = \bW^\transpose \bLambda \, \bW
\ee
where $\bLambda$ is diagonal.
The quantities $\bW \hat\theta$ are uncorrelated
because their covariance matrix is diagonal
\be
  \bW \langle \Delta\hat \theta \Delta\hat
    \theta^\transpose \rangle \bW^\transpose
  = \bLambda^{-1}
  \ .
\ee
Being inverse variances of $\bW \hat\theta$,
the diagonal elements of the diagonal matrix $\bLambda$
are necessarily positive.
The Fisher matrix of the decorrelated quantities $\bW \hat\theta$ is
the diagonal matrix $\bLambda$, since
\be
  \left( \bW
    \langle \Delta\hat \theta \Delta\hat \theta^\transpose \rangle
    \bW^\transpose \right)^{-1}
  = \bLambda
  \ .
\ee

Without loss of generality,
the decorrelated quantities $\bW \hat\theta$ can be scaled to unit variance
by multiplying them by the square root of the corresponding diagonal
element of $\bLambda$.
Scaled to unit variance, the decorrelation matrices satisfy
\be
\label{decorr1}
  \bF = \bW^\transpose \bW
  \ .
\ee
There are infinitely many distinct decorrelation matrices
satisfying equation~(\ref{decorr1}).
Any orthogonal rotation $\bO$ of a decorrelation matrix $\bW$
yields another decorrelation matrix $\bV = \bO \bW$,
since
$\bV^\transpose \bV = \bW^\transpose \bO^\transpose \bO \bW
= \bW^\transpose \bW = \bF$.
Conversely,
if $\bV$ and $\bW$ are two decorrelation matrices
satisfying
$\bF = \bV^\transpose \bV = \bW^\transpose \bW$,
then $\bV = \bO \bW$
is an orthogonal rotation of $\bW$,
since
$(\bW^\transpose)^{-1} \bV^\transpose \bV \bW^{-1} =
(\bV \bW^{-1})^\transpose \bV \bW^{-1} = \1$
shows that $\bO = \bV \bW^{-1}$ is an orthogonal matrix.

\subsection{Prewhitened power spectrum}

The prewhitened nonlinear power spectrum $X(k)$ is the Fourier transform
of the prewhitened nonlinear correlation function $X(r)$
defined in terms of the nonlinear correlation function $\xi(r)$ by
(Paper~3 \S5)
\be
  X(r) \equiv {2 \, \xi(r) \over 1 + [1 + \xi(r)]^{1/2}}
  \ .
\ee
In this paper, as in Paper~3,
hats are used to denote estimators, so that
$\hat X(k)$ with a hat on denotes an estimate of
the prewhitened power spectrum measured from a galaxy survey
(Paper~3 \S7).
The quantity $X(k)$ without a hat denotes
the prior prewhitened power spectrum,
the model power spectrum whose viability is tested in a likelihood analysis.

The covariance of the prewhitened nonlinear power spectrum is
approximately equal to the inverse of its Fisher matrix,
denoted $E^{\alpha\beta}$ (Paper~3 \S6),
\be
  \langle \Delta X_\alpha \Delta X_\beta \rangle^{-1}
  =
  E^{\alpha\beta}
  \ .
\ee

The weighting of data that yields the best (minimum variance)
estimate of power depends on the choice of prior power,
so that the estimated prewhitened power $\hat X(k)$ depends (weakly) on
the prior prewhitened power $X(k)$.
An incorrect guess for the prior will not bias the
estimator high or low, but merely makes it slightly noisier than 
the minimum allowed by the Fisher matrix (Tegmark et al.\ 1997).
The Maximum Likelihood solution to the prewhitened power spectrum
of a galaxy survey can be obtained by folding the estimate $\hat X(k)$
back into the prior $X(k)$ and iterating to convergence
(Tegmark et al.\ 1998).

In this paper, the prior power $X(k)$ is simply treated as some fiducial
prewhitened power spectrum.
In all examples illustrated, the prior power spectrum is taken to be an
observationally concordant $\Lambda$CDM model from Eisenstein \& Hu (1998).

\subsection{Band-powers}

It is desirable to define estimated
band-powers $\hat B(k_\alpha)$ and band-power windows
$W(k_\alpha,k_\beta)$ for the estimated prewhitened power spectrum
$\hat X(k)$ in a physically sensible way.
One of the problems is that the prewhitened power spectrum
is liable to vary by orders of magnitude,
so that, carelessly defined,
a windowed power could be dominated by power leaking in from
the wings of the band,
rather than being a sensible average of power in the band.

Let band-powers $\hat B(k_\alpha)$ be defined by
\be
\label{Bgk}
  {\hat B(k_\alpha) \over \chi(k_\alpha)} \equiv
    \int_0^\infty W(k_\alpha,k_\beta) {\hat X(k_\beta) \over \chi(k_\beta)}
    \, {4\PI k_\beta^2 \dd k_\beta \over (2\PI)^3}
\ee
where $\chi(k)$ is some scaling function, to be chosen below
in equation~(\ref{gX}).
In manipulations, it can be convenient to treat $\chi$ as a matrix
that is diagonal in Fourier space with diagonal entries $\chi(k)$.
The band-power windows $W(k_\alpha,k_\beta)$ in equation~(\ref{Bgk})
are normalized to unit integral
\be
\label{intW}
  \int W(k_\alpha,k_\beta)
    {4 \PI (k_\alpha k_\beta)^{3/2} \over (2\PI)^3}
    \, \dd \ln k_\beta
  = 1
  \ ,
\ee
the slightly complicated form of which is chosen so as to make its
discretized equivalent, equation~(\ref{sumW}), look simple.
Continuous matrices must be discretized in order to
manipulate them numerically.
As described in \S2.3 of Paper~3,
discretization should be done in such a way as to preserve the
inner product in Hilbert space.
Discretized on a logarithmic grid of wavenumbers,
a continuous vector such as $\hat X(k_\alpha)$ becomes
the discrete vector
$\hat \bX_{k_\alpha} \equiv \hat X(k_\alpha)$\discretionary{}{}{}$
[4 \PI k_\alpha^3 \Delta \ln k / (2\PI)^3]^{1/2}$,
while a continuous matrix such as $W(k_\alpha,k_\beta)$
becomes the discrete matrix
$\bW_{k_\alpha k_\beta} \equiv
W(k_\alpha,k_\beta)
\,$\discretionary{}{}{}$4\PI (k_\alpha k_\beta)^{3/2}$\discretionary{}{}{}$\Delta\ln k/(2\PI)^3$.
Thus, discretized on a logarithmic grid of wavenumbers, equation~(\ref{Bgk})
translates into an equation for the discrete band-powers
$\hat \bB_{k_\alpha} \equiv \hat B(k_\alpha)$\discretionary{}{}{}$
[4 \PI k_\alpha^3 \Delta \ln k / (2\PI)^3]^{1/2}$,
\be
\label{Bg}
  {\hat \bB_{k_\alpha} \over \chi(k_\alpha)}
  = \sum_{k_\beta} \bW_{k_\alpha k_\beta}
    {\hat\bX_{k_\beta} \over \chi(k_\beta)}
  \ .
\ee
Each row of the discrete matrix $\bW_{k_\alpha k_\beta}$ represents
a band-power window for the discrete band-power $\hat \bB_{k_\alpha}$,
and it is these discrete band-power windows $\bW_{k_\alpha k_\beta}$
that are plotted in Figures~\ref{Xpcd_sdss}--\ref{Fhg}.
Normalizing each discrete band-power window $\bW_{k_\alpha k_\beta}$
to unit sum
\be
\label{sumW}
  \sum_{k_\beta} \bW_{k_\alpha k_\beta}
  = 1
\ee
ensures that the band-power $\hat \bB_{k_\alpha}$
represents an average of prewhitened power in the band.

The scaling function $\chi(k)$
in equations~(\ref{Bgk}) or (\ref{Bg})
is introduced to ensure a physically sensible definition
of the band-power windows.
Rewriting equation~(\ref{Bg}) as
\be
  \hat \bB_{k_\alpha}
  = \sum_{k_\beta} \chi(k_\alpha) \bW_{k_\alpha k_\beta} \chi(k_\beta)^{-1}
    \, \hat\bX_{k_\beta}
\ee
makes it plain that choosing different scaling functions $\chi$
is equivalent to rescaling the band-power windows as
$\bW \rightarrow \chi \bW \chi^{-1}$.

Suppose for example that the scaling function in equation~(\ref{Bg})
were chosen to be one, $\chi(k) = 1$,
so that $\hat\bB = \bW \hat\bX$.
Then a band-power $\hat\bB_{k_\alpha}$ could be dominated by
power leaking in from wavenumbers $k_\beta$ where $\hat\bX_{k_\beta}$ is large,
even though the window $\bW_{k_\alpha k_\beta}$ were small there,
which is not good.

A better choice, the one adopted in this paper,
is to set the scaling function $\chi(k)$
equal to the discretized prior prewhitened power $\bX_k$,
\be
\label{gX}
  \chi(k) = \bX_{k}
  \ ,
\ee
so that the discrete band-powers $\hat\bB_{k_\alpha}$,
equation~(\ref{Bg}), are defined by
\be
\label{B}
  {\hat \bB_{k_\alpha} \over \bX_{k_\alpha}}
  = \sum_{k_\beta} \bW_{k_\alpha k_\beta}
    {\hat\bX_{k_\beta} \over \bX_{k_\beta}}
  \ .
\ee
Since the expectation is that
$\hat\bX_{k_\beta} / \bX_{k_\beta} \approx 1$,
the definition~(\ref{B}) ensures that the contribution
from power $\hat\bX_{k_\beta}$ at wavenumber $k_\beta$
to a band-power $\hat \bB_{k_\alpha}$
is large where the window $\bW_{k_\alpha k_\beta}$ is large,
and small where the window is small, as is desirable.

\subsection{Decorrelated band-powers}

Equation~(\ref{B}) expresses the estimated band-powers $\hat\bB$
as linear combinations of scaled prewhitened powers $\hat\bX/\bX$.
The discrete Fisher matrix $\bG$ of the scaled prewhitened power is
(again, it is convenient to treat the scaling function $\bX_k$
as a matrix that is diagonal in Fourier space, with diagonal entries $\bX_k$)
\be
\label{G}
  \bG
  = \bX \langle \Delta\hat\bX \, \Delta\hat\bX^\transpose \rangle^{-1} \bX
  = \bX \bE \bX
\ee
where
$\bE_{k_\alpha k_\beta} \equiv
E(k_\alpha,k_\beta)
\,$\discretionary{}{}{}$4\PI (k_\alpha k_\beta)^{3/2}$\discretionary{}{}{}$\Delta\ln k/(2\PI)^3$
is the discretized Fisher matrix of the prewhitened power
(Paper~3 \S6).

All the band-power windows $\bW$ constructed in this paper
are decorrelation matrices, satisfying
\be
\label{GW}
  \bG
  = \bW^\transpose \bLambda \, \bW
\ee
where $\bLambda$ is diagonal in Fourier space.
By construction, the estimated scaled band-powers
$\hat\bB/\bX = \bW \hat \bX/\bX$ are uncorrelated,
i.e.\ their covariance matrix is diagonal in Fourier space
\be
  \bX^{-1} \langle \Delta\hat\bB \, \Delta\hat\bB^\transpose \rangle \bX^{-1}
  = \bLambda^{-1}
  \ .
\ee
The Fisher matrix of the decorrelated scaled band-powers $\hat\bB/\bX$
is the diagonal matrix
\be
\label{FXB}
  \bX \langle \Delta \hat\bB \, \Delta \hat\bB^\transpose \rangle^{-1} \bX
  = \bLambda
  \ .
\ee

The decorrelation process
decorrelates not only the scaled prewhitened power $\hat\bX/\bX$,
but also the prewhitened power spectrum $\hat\bX$ itself,
as is to be expected since the scaling factors $1/\bX$ are just constants.
While the band-power windows for the scaled prewhitened power are $\bW$,
the band-power windows for the prewhitened power itself are $\bX \bW \bX^{-1}$,
since $\hat\bB = (\bX \bW \bX^{-1})\hat \bX$.
The band-power windows $\bX \bW \bX^{-1}$ are themselves decorrelation matrices,
equation~(\ref{decorr}),
for the prewhitened power $\hat\bX$, satisfying
\be
  \bE =
    ( \bX \bW \bX^{-1} )^\transpose
    ( \bX^{-1} \bLambda \, \bX^{-1} )
    ( \bX \bW \bX^{-1} )
  \ .
\ee
The decorrelated band-powers
$\hat\bB = (\bX \bW \bX^{-1}) \hat \bX$,
are uncorrelated because their covariance matrix is diagonal
in Fourier space
(recall again that the scaling function $\bX$ is effectively
a diagonal matrix in Fourier space)
\be
  \langle \Delta\hat\bB \, \Delta\hat\bB^\transpose \rangle
  = 
    ( \bX \bW \bX^{-1} )
    \langle \Delta \hat\bX \, \Delta \hat\bX^\transpose \rangle
    ( \bX \bW \bX^{-1} )^\transpose
  = \bX \, \bLambda^{-1} \bX
  \ .
\ee
The Fisher matrix of the decorrelated band-powers $\hat\bB$
is the diagonal matrix
\be
\label{FB}
  \langle \Delta \hat\bB \, \Delta \hat\bB^\transpose \rangle^{-1}
  = \bX^{-1} \bLambda \, \bX^{-1}
  \ .
\ee

\Ffig

\subsection{Interpretation of scaled powers as log powers}

It is interesting, although peripheral to the central thread of this paper,
to note that the scaled powers $\hat\bB/\bX$ and $\hat\bX/\bX$
can be interpreted in terms of log powers,
at least in the limit of a large quantity of data,
where $\Delta\hat\bX \equiv \hat\bX-\bX \ll \bX$.

In this limit
$(\hat\bB/\bX) - 1 \approx \ln(\hat\bB/\bX)$
and
$(\hat\bX/\bX) - 1 \approx \ln(\hat\bX/\bX)$.
Thus equation~(\ref{B}), with one subtracted from both sides,
can be rewritten
\be
  \ln\left( {\hat\bB_{k_\alpha} \over \bX_{k_\alpha}} \right)
  = \sum_{k_\beta} \bW_{k_\alpha k_\beta}
    \ln\left( {\hat\bX_{k_\beta} \over \bX_{k_\beta}} \right)
\ee
or equivalently
\be
\label{lnB}
  \ln \hat\bB_{k_\alpha}
  = C_{k_\alpha} + \sum_{k_\beta} \bW_{k_\alpha k_\beta}
    \ln \hat\bX_{k_\beta}
\ee
where
$C_{k_\alpha} \equiv
\sum_{k_\beta} \bW_{k_\alpha k_\beta} \ln(\bX_{k_\alpha}/\bX_{k_\beta})$
are constants that are zero if the prior prewhitened power $\bX_k$
happens to vary as a power law with wavenumber $k$,
and in practice should be close to zero as long as the band-power windows
are narrow.

Equation~(\ref{lnB}) shows that,
modulo the small constant offsets $C$,
the log band-powers $\ln\hat\bB$ can be regarded as windowed averages
of the log prewhitened powers $\ln\hat\bX$.
Irrespective of the offsets $C$,
the Fisher matrix of the log prewhitened powers
approximates the Fisher matrix $\bG$, equation~(\ref{G}),
\be
  \langle \Delta \ln\hat\bX \, \Delta \ln\hat\bX^\transpose \rangle^{-1}
  \approx
  \bX \langle \Delta\hat\bX \, \Delta\hat\bX^\transpose \rangle^{-1} \bX
  = \bG
\ee
and the decorrelation matrices $\bW$, equation~(\ref{GW}),
can be regarded as decorrelation matrices for the
log prewhitened power $\ln\hat\bX$.

\subsection{Fisher matrices of power in SDSS and LCRS}

Constructing decorrelated band-powers requires knowing the Fisher matrix.
Figure~\ref{F}
shows the discrete Fisher matrices $\bG$, equation~(\ref{G}),
of the scaled prewhitened nonlinear power spectra $\hat\bX/\bX$
of SDSS and LCRS.
The Fisher matrix was computed in the FKP approximation,
as described in \S6 of Paper~3.

The prior power spectrum $X(k)$ is taken to be a
$\Lambda$CDM model of Eisenstein \& Hu (1998),
with observationally concordant parameters
$\Omega_\Lambda = 0.7$, $\Omega_m = 0.3$,
$\Omega_b h^2 = 0.02$,
and $h \equiv H_0 / (100 \, \km \, {\rmn s}^{-1} \Mpc^{-1}) = 0.65$,
nonlinearly evolved according to the Peacock \& Dodds (1996) formula.
$N$-body simulations by Meiksin, White \& Peacock (1999)
indicate that nonlinear evolution tends to suppress baryonic wiggles
in the (unprewhitened) power spectrum,
whereas the Peacock \& Dodds transformation preserves the wiggles.
For simplicity we retain the Peacock \& Dodds formalism,
notwithstanding its possible defects.

The FKP approximation is valid at wavelengths small compared to
the scale of the survey,
so should work better at smaller scales in surveys with broad
contiguous sky coverage,
and worse at larger scales in slice or pencil beam surveys.
Thus the FKP approximation should work reasonably well
at all but the largest scales in SDSS,
but probably fails in LCRS at intermediate and large scales.
We suspect that the FKP approximation
is liable to underestimate the Fisher information
(overestimate the error bars) in LCRS,
since the density in a thin slice is correlated with
(and hence contains information about) the density outside the survey volume.
A reliable assessment of the extent to which the FKP approximation
under- or over-estimates information awaits future explicit calculations,
such as described in Tegmark et al.\ (1998).

At the median depth $\approx 300 \, h^{-1} \Mpc$ of the LCRS survey,
each $1\fdg5$ slice is $7.5 \, h^{-1} \Mpc$ thick,
so the FKP approximation might be expected to break down
in LCRS at wavenumbers
$k \la \PI/7.5 \  h \, \Mpc^{-1} = 0.4 \, h \, \Mpc^{-1}$.

The Fisher matrix shown in Figure~\ref{F} contains negative elements,
notably at intermediate wavenumbers $k \sim 1 \, h \, \Mpc^{-1}$,
whereas for Gaussian fluctuations the Fisher matrix would be
everywhere positive.
According to equation~(32) of Paper~3,
the Fisher matrix of the power spectrum for Gaussian fluctuations would be
\be
  F^{\alpha\beta}
  = \frac{1}{2} \, D^\alpha_{ij} C^{-1 ik} C^{-1 jl} D^\beta_{kl}
  \ .
\ee
Thus in Fourier space the elements of the Fisher matrix of the power spectrum
\be
\label{Fkkgau}
  F(k_\alpha, k_\beta)
  = \frac{1}{2} \int \left| C^{-1}\!(k_\alpha \n_i, k_\beta \n_j) \right|^2
    {\dd\Omega_i \, \dd\Omega_j \over (4\PI)^2}
\ee
would all be necessarily positive.
The quantities $\dd\Omega$ in equation~(\ref{Fkkgau}) are intervals of solid angle,
and the integration is over all unit directions $\n$ of the wavevectors.
This positivity of the Fisher matrix for Gaussian fluctuations
is separate from and in addition to the fact that the Fisher
matrix is positive definite,
i.e.\ has all positive eigenvalues.

The fact that some elements of the Fisher matrix in Figure~\ref{F} are
negative is a consequence of nonlinearity.
Nonlinear evolution induces broad positive correlations
in the covariance matrix of estimates of the power spectrum (Paper~3 Fig.~2),
leading to negative elements in the Fisher matrix,
the inverse of the covariance of power.
Prewhitening the nonlinear power spectrum substantially narrows
the covariance of power, and diminishes the amount of negativity
in the Fisher matrix, but negative elements remain.

\section{Example Band-Powers}

\subsection{Principal Component Decomposition}
\label{PCD}

Let $\bO$ be an orthogonal matrix that diagonalizes the Fisher matrix
$\bG$ of the scaled prewhitened power $\hat\bX/\bX$
\be
  \bG = \bO^\transpose \bLambda \, \bO
  \ .
\ee
Then $\bO$ is a decorrelation matrix.
The decorrelated band powers $\bO \hat\bX/\bX$
constitute the principal component decomposition of the
scaled prewhitened power spectrum.

\Xpcdsdssfig

Figure~\ref{Xpcd_sdss}
shows the band-power windows,
the rows of $\bO$,
for the principal component decomposition of the scaled prewhitened power
spectrum of SDSS.
Since the Fisher matrix $\bG$ is already fairly narrow in Fourier space,
Figure~\ref{F},
one might have thought that its eigenmodes would be similarly narrow
in Fourier space.
But this is not so.
In fact the eigenmodes, the band-power windows, are generally broad and,
aside from the first, fundamental mode, generally wiggly.
This makes the principal component decomposition of the power spectrum
of little practical use, as previously concluded in Paper~2.
This should of course not be confused with
a principal component decomposition of the density
field itself, which can be of great utility
(Vogeley \& Szalay 1996; Tegmark et al.\ 1998).

Elsewhere in this paper plotted band-power windows
are scaled to unit sum over the window, equation~(\ref{sumW}),
but in Figure~\ref{Xpcd_sdss}
the band-power windows $\bO$ are scaled instead to unit sum of squares
\be
\label{sumO}
  \sum_{k_\beta} \bO_{k_\alpha k_\beta}^2
  = 1
  \ .
\ee
In other words, the band-power windows plotted in Figure~\ref{Xpcd_sdss}
are just the orthonormal eigenfunctions of the Fisher matrix $\bG$.
The problem with scaling the band-power windows to unit sum is
that it makes the wigglier windows oscillate more fiercely,
obscuring the less wiggly windows.
However, the labelling of the band-power windows in Figure~\ref{Xpcd_sdss}
is in order of the expected variances of the band-powers
normalized to unit sum over the window,
$\sum_{k_\beta} \bO_{k_\alpha k_\beta} = 1$,
not to unit sum of squares.

\subsection{Cholesky Decomposition}
\label{cholesky}

Clearly it is desirable to use the infinite freedom of choice
(\S\ref{decorrelationW}) of decorrelation matrices
to engineer band-power windows that are narrow and not wiggly.
To understand why principal component decomposition works badly,
consider the mathematical theorem that the eigenmodes of a real symmetric matrix
are unique up to arbitrary orthogonal rotations amongst degenerate eigenmodes,
that is, amongst eigenmodes with the same eigenvalue.
In the present case the eigenvalues of the Fisher matrix,
even if not degenerate, are nevertheless many and finely spaced,
and there is much potential for nearly degenerate eigenmodes to
mix in random unpleasant ways.
This suggests that mixing can be reduced in Fourier space
by lifting the degeneracy of eigenvalues in Fourier space.

One way to lift the degeneracy is to
multiply the rows and columns of the Fisher matrix by a strongly varying
function $\gamma(k)$ of wavenumber $k$ before diagonalizing.
That is,
consider diagonalizing the scaled Fisher matrix
$\gamma^\transpose \bG \, \gamma$
\be
  \gamma^\transpose \bG \, \gamma = \bO^\transpose \bLambda \, \bO
\ee
where $\gamma$ is some scaling matrix,
$\bO$ is orthogonal, and $\bLambda$ is diagonal.
Then $\bW = \bO \gamma^{-1}$ is a decorrelation matrix,
with $\bG = \bW^\transpose \bLambda \bW$.
Now take the scaling matrix $\gamma$ to be a diagonal matrix in Fourier space
with diagonal entries $\gamma(k)$,
and let $\gamma(k)$ be strongly varying with $k$.

In the limit of an infinitely steeply varying scaling function,
$\gamma(k_1) \gg \gamma(k_2) \gg \ldots $,
the resulting decorrelation matrix $\bW$ becomes upper triangular,
as argued in \S5 of Paper~2.
As pointed out by Tegmark \& Hamilton (1998)
and Tegmark (1997a),
this choice of decorrelation matrix is just a
Cholesky decomposition of the Fisher matrix
\be
  \bG = \bU^\transpose \bU
\ee
where $\bU$ is upper triangular.

If the scaling function $\gamma$ is chosen to be infinitely steep
in the opposite direction,
$\gamma(k_1) \ll \gamma(k_2) \ll \ldots $,
then the resulting decorrelation matrix is lower triangular.
This is equivalent to a Cholesky decomposition of the Fisher matrix
\be
  \bG = \bL^\transpose \bL
\ee
with $\bL$ lower triangular.

Such a Cholesky decomposition has been successfully employed
to construct a decorrelated power spectrum of the CMB
from both the COBE DMR data (Tegmark \& Hamilton 1998) and
the 3 year Saskatoon data (Knox, Bond \& Jaffe 1998).

\Xcholfig

Figure~\ref{Xchol}
shows band-power windows from the upper triangular Cholesky decomposition
of the Fisher matrices $\bG$ of the scaled prewhitened power
of SDSS and LCRS.
These band-powers are a considerable improvement over
the principal component decomposition of Figure~\ref{Xpcd_sdss},
in the sense that they are narrower and less wiggly.
However,
the Cholesky decomposition has two defects.
The first defect is that the band-power windows are skewed.
By construction, the upper triangular Cholesky windows vanish
to the left of the diagonal element.
Figure~\ref{Xchol} shows that the Cholesky windows have a tail to the right
that, although small, is nevertheless large enough to start to become worrying,
especially in LCRS.
A lower triangular Cholesky decomposition leads to band-power windows
skewed in the opposite direction (not plotted).

The second defect of the Cholesky decomposition is that
it does not tolerate negative eigenvalues in the Fisher matrix well,
a point previously discussed in Paper~2.
In Figure~\ref{Xchol},
the Fisher matrices have been truncated to a maximum wavenumber
of $k_{\max} = 0.015 \, h \, \Mpc^{-1}$ in SDSS,
and $k_{\max} = 0.034 \, h \, \Mpc^{-1}$ in LCRS,
to ensure that the computed Fisher matrix has no negative eigenvalues.
Although the Fisher matrix should in theory be positive definite,
meaning that all its eigenvalues should be positive,
numerically it acquires negative eigenvalues
when the resolution in $k$ is increased
to the point where the spacing $\Delta k$ between adjacent wavenumbers
is less than of the order of the inverse scale size of the survey.
Presumably negative eigenvalues appear
thanks to a combination of numerical noise and the various approximations
that go into evaluating the Fisher matrix.
The band-power windows associated with negative eigenvalues are ill-behaved,
unlike those shown in Figure~\ref{Xchol}.
The ill-behaved band-power windows tend to infect their neighbours,
and as more and more eigenvalues become negative with increasing resolution,
the whole system of Cholesky windows devolves into chaos.

Replacing the Fisher matrix with a version of it
in which negative eigenvalues are replaced by zero
does not solve the difficulty.

It is possible to extend the band-powers to smaller wavenumbers
by using a linear instead of logarithmic binning in wavenumber.
With linear binning, the maximum wavenumber is
$k_{\max} = \PI/1024 \  h \, \Mpc^{-1}$ in SDSS,
$k_{\max} = \PI/640 \  h \, \Mpc^{-1}$ in LCRS.

\subsection{Square Root of the Fisher Matrix}
\label{sqrt}

While the Cholesky band-powers are a definite improvement
over the principal component decomposition,
evidently one would like the band-power windows to be more symmetric
about their peaks.
One way to achieve this is to choose the decorrelation matrix
to be symmetric,
which corresponds to choosing it to be the
square root of the Fisher matrix
\be
  \bG = \bG^{1/2} \bG^{1/2}
  \ .
\ee
The band-powers $\bG^{1/2} \hat\bX/\bX$ are uncorrelated,
with unit variance.
Tegmark \& Hamilton (1998)
previously used the square root of the Fisher matrix
to construct the decorrelated power spectrum
of CMB fluctuations from the {\it COBE\/} data, and found
the resulting window functions to be narrow, non-negative and
approximately symmetric. One might therefore hope that these
properties would hold also for the case of galaxy surveys.

\Fhfig

Figure~\ref{Fh}
shows the band-power windows constructed from the square root
of the Fisher matrix of the scaled prewhitened power of SDSS and LCRS.
The plotted windows, the rows of $\bG^{1/2}$,
are normalized to unit sum as in equation~(\ref{sumW}).
The band-power windows are all nicely narrow and symmetrical.
The Fisher matrix $\bG$ itself is already fairly narrow about
the diagonal, Figure~\ref{F},
and taking its square root narrows it even further,
in much the same way that taking the square root of a Gaussian matrix
$M_{x x'} = e^{-(x-x')^2}$ narrows it by $2^{1/2}$,
to $M_{x x'}^{1/2} = e^{-2(x-x')^2}$.

In practice, the square root of the Fisher matrix is constructed
by diagonalizing the Fisher matrix,
$\bG = \bO^\transpose \bLambda \bO$, and setting
\be
\label{sqrtG}
  \bG^{1/2} = \bO^\transpose \bLambda^{1/2} \bO
\ee
the positive square root of the eigenvalues being taken.
Negative eigenvalues,
which as remarked in \S\ref{cholesky}
presumably arise from a combination of numerical
noise and the approximations that go into evaluating the Fisher matrix,
must be replaced by zero in equation~(\ref{sqrtG}).
The negative eigenvalues are invariably small in absolute value
compared to the largest positive eigenvalue.

Replacing negative eigenvalues in equation~(\ref{sqrtG}) by zeros
causes the band-power windows, the rows of $\bG^{1/2}$,
to become linearly dependent,
and the resulting band-powers to become correlated.
Remarkably enough, however, it is still possible to regard the band-powers
$\bG^{1/2} \hat\bX/\bX$ as remaining uncorrelated with unit variance.
The negative eigenvalues correspond to noisy eigenmodes of the Fisher matrix.
Replacing the negative eigenvalues by zero in equation~(\ref{sqrtG})
corresponds to eliminating the contribution of these noisy modes
from the band-power windows $\bG^{1/2}$.
Now if the Fisher matrix were calculated perfectly,
then it would have no negative eigenvalues,
so correctly the noisy modes would make some positive contribution
to $\bG^{1/2}$.
The contribution would however be small because the modes are noisy
and the corresponding eigenvalues are small.
Thus one should imagine that the $\bG^{1/2}$ that results
from setting negative eigenvalues to zero differs only slightly
from the true $\bG^{1/2}$
that yields band powers $\bG^{1/2} \hat\bX/\bX$ with unit covariance matrix.

As the resolution of the Fisher matrix is increased,
more and more eigenvalues become negative.
The effect on the band-power windows is intriguing and enlightening.
If for example the resolution is doubled,
then there are twice as many band-powers.
The old band-power windows retain essentially the same shapes as before,
except that they are computed with twice as much resolution.
The new band-power windows interleave with the old ones,
and have shapes that vary smoothly between the shapes of the
adjacent band-powers.
Normalized to unit sum over the window,
the variance of each band-power doubles,
in just such a fashion that the net information content,
the summed inverse variance, remains constant.
This behaviour persists to the highest resolution that we have tested it,
$\Delta\log k = 1/1024$.


It is worth emphasizing just how remarkable
this behaviour of the band-powers is.
It seems as though the decorrelated band-power windows are converging,
as the resolution goes to infinity, towards a well-defined continuous shape.
Yet, in spite of the asymptotic constancy of shape,
the variance associated with each window seems to 
continue increasing inversely with resolution,
i.e.\ inversely with the number of windows,
tending to infinity in the continuous limit
in such a fashion that the net variance over any fixed interval of wavenumber
remains constant.

How can this be?
How can there be infinitely many 
uncorrelated band-power estimates?
It would seem that as the resolution increases,
the band-power windows manage to remain decorrelated by incorporating
into themselves small contributions of noisy modes that increase the variance
of the band-power while changing the windows only slightly.
In the continuum limit, the variance of each band-power would continue to
increase indefinitely while the shape of the window changes only
infinitesimally.

In practice,
the band-power windows at the largest scales,
nominal wavenumbers less than the natural resolution of the survey,
become jittery at high resolution.
The `natural resolution' here is defined empirically,
as the highest resolution for which all the eigenvalues of the Fisher matrix
remain numerically positive,
$\Delta k = \PI/1024 \  h \, \Mpc^{-1}$ in SDSS,
and $\Delta k = \PI/640 \  h \, \Mpc^{-1}$ in LCRS.
We attribute the jitter in part to the fact that the pair integrals
$R(r;\mu)$ used to compute the Fisher matrix are themselves
only measured with finite resolution and accuracy,
and in part to the fact that there is practically no signal
at the largest scales, so all the relevant eigenvalues are small,
and the numerics have a harder time distinguishing signal from noise.
Perhaps there is a better algorithm than the simple one adopted here
of setting negative eigenvalues to zero,
but the simple algorithm does seem to work well enough.

The situation is illustrated in Figure~\ref{Fh},
which shows the first band-power window in the sequence for SDSS,
the one nominally corresponding to $k = 0.001 \, h \, \Mpc^{-1}$,
multiplied by a factor of 10 to show it more clearly.
This nominal wavenumber is smaller by a factor of $\sim 3$ than
the smallest measurable wavenumber $k \sim 0.003 \, h \, \Mpc^{-1}$ in SDSS,
and the computed band-power accordingly retreats to a
larger effective wavenumber, where there is signal.
At scales comparable to the natural resolution of the survey,
$k \sim \Delta k \sim 0.003 \, h \, \Mpc^{-1}$,
the resolution $\Delta\log k = 1/32$ used in Figure~\ref{Fh}
is 10 times higher than the natural resolution.
For comparison,
Figure~\ref{Fh} also shows (shaded) the same band-power computed at a
resolution of $\Delta\log k = 1/1024$,
several hundred times the natural resolution.
At this high resolution, the band-power oscillates finely,
but overall remains under control.
Such robust behaviour contrasts with the chaotic behaviour
of the Cholesky windows when similarly pushed beyond the natural resolution
of the survey.

\Fhlinfig

The methods of Paper~3 permit the Fisher matrix to be
discretized over any arbitrary grid of wavenumbers
(although all the explicit examples in Paper~3 employ a logarithmic grid).
Figure~\ref{Fh_lin}
shows the band-power windows constructed from the square root
of the Fisher matrix of the scaled prewhitened power of SDSS
for a linearly spaced rather than logarithmically spaced grid of wavenumbers.
These band-power windows have essentially the same shapes
as those computed on the logarithmic grid, Figure~\ref{Fh}.
The difference is in the number of band-power windows and in the
resolution with which they are defined.

Figure~\ref{Fh_lin} shows, more clearly than Figure~\ref{Fh},
that the band-powers are broader for LCRS than for SDSS,
reflecting the smaller effective volume,
hence lower resolution in Fourier space, of LCRS.
The slice geometry of LCRS leads to wings on the band-powers.
The wings are fairly mild here,
but in pencil-beam surveys the wings can become quite broad,
potentially leading to significant aliasing between power at small
and large wavenumbers
(Kaiser \& Peacock 1991).

For LCRS, the first band-power plotted in Figure~\ref{Fh_lin},
at a nominal wavenumber of $k = \PI/1024 \  h \, \Mpc^{-1}$, is jittery,
illustrating again that band-powers at nominal wavenumbers
less than the natural resolution of the survey,
$k = \PI/640 \  h \, \Mpc^{-1}$ for LCRS,
tend to become jittery at high resolution.
Again, we attribute this in part to the fact that the pair integrals
$R(r;\mu)$ used to compute the Fisher matrix are themselves computed
with finite accuracy
(in LCRS, the pair integrals were evaluated by the Monte Carlo method),
and in part to the fact that the numerics have a harder time
distinguishing signal from noise when the signal is weak.

\subsection{Scaled Square Root of the Fisher Matrix}

Notwithstanding the success of the square root of the Fisher matrix
in many cases, it does not work perfectly in all cases.
The problem is that the symmetry of $\bG^{1/2}$
does not imply symmetry of the band-power windows,
the rows of $\bG^{1/2}$, about their diagonals.
If the Fisher matrix $\bG$ varies steeply along the diagonal,
then the band-power windows can turn out quite skewed.
In practice, this happens for example in the case of a perfect,
shot-noiseless survey,
where the information contained in the power spectrum increases without limit
as $k \rightarrow \infty$.
The case of a noiseless survey was applied in \S9.3 of Paper~3
to compute the effective FKP constants $\mu(k)$
to be used in an FKP pair-weighting
when measuring the prewhitened nonlinear power spectrum of a survey.

\Fhgfig

The difficulty can be remedied by scaling
the Fisher matrix before taking its square root.
Let $\gamma$ be any scaling matrix.
Then
\be
\label{Wgg}
  \bW = ( \gamma^\transpose \bG \, \gamma )^{1/2} \gamma^{-1}
\ee
is a decorrelation matrix, satisfying $\bG = \bW^\transpose \bW$.

In the case of the perfect, noiseless survey,
a diagonal scaling matrix $\gamma$ with diagonal elements
\be
\label{gscale}
  \gamma(k) = k^{3/2}
\ee
proves empirically to work well.
Figure~\ref{Fhg}
compares band-power windows
obtained from the scaled versus unscaled square root of the Fisher matrix,
for the perfect, noiseless survey.
For clarity
only a single band-power window,
the one centred at $k = 1 \, h \, \Mpc^{-1}$,
is shown, but this window is representative.
Scaling the Fisher matrix with $\gamma(k) = k^{3/2}$ before taking its
square root in this case helps to rectify the skew in the band-power windows
from the unscaled square root $\bG^{1/2}$.
The results shown in Figure~11 of Paper~3
were obtained using the decorrelation matrix of equation~(\ref{Wgg})
scaled by the scaling function $\gamma(k) = k^{3/2}$, equation~(\ref{gscale}).

\xiafig

\section{Decorrelated Power Spectrum}
\label{power}

The goal of this paper has been to show how to obtain uncorrelated
band-powers $\hat\bB$ that can be plotted, with error bars, on a graph.
In \S\ref{sqrt} it was found that,
if the band-powers are decorrelated with the square root of the Fisher matrix,
then the band-powers can be computed at as high (or low) a resolution
as one cares,
and the band-powers will remain effectively uncorrelated.
The higher the resolution, the larger the error bars on the band-powers.
The same result holds if the band-powers are decorrelated with
the scaled square root of the Fisher matrix, equation~(\ref{Wgg}).

If the band-power windows $\bW$ are scaled to unit sum,
equation~(\ref{sumW}),
then the Fisher matrix of the scaled band-powers $\hat\bB/\bX$
is the diagonal matrix $\bLambda$
in $\bG = \bW^\transpose \bLambda \bW$,
equation~(\ref{FXB}).
The quantity that remains invariant with respect to resolution
(and linear or logarithmic binning) is the inverse variance,
also called the information,
per unit log wavenumber $\dd I/\dd \ln k$
\be
\label{dIdk}
  {\dd I \over \dd \ln k} = {\Lambda_k \over \Delta \ln k}
\ee
where $\Lambda_k$ is the diagonal element of $\bLambda$ associated with the
scaled band-power $\hat\bB_k/\bX_k$,
and $\Delta\ln k$ is the resolution of the matrix at wavenumber $k$.

The association of a band-power $\hat\bB_k$ with wavenumber $k$
relies on the band-power window being narrow about that wavenumber.
In practice the band-power windows have a finite width,
and the correspondence with wave number is not exact,
a reflection of the uncertainty principle.
As discussed in \S\ref{sqrt},
the decorrelated band-power windows would remain of finite width
even if they were resolved at infinite resolution.

\xikfig

Figure~\ref{xia} shows the information per unit log wavenumber
$\dd I/\dd \ln k$
in the scaled prewhitened power spectra of
SDSS, LCRS, and also the {\it IRAS\/} 1.2~Jy survey,
for the $\Lambda$CDM prior power spectrum.
These information curves are similar to those presented by Tegmark (1997b),
but are more accurate, and valid also in the nonlinear regime.

It should be cautioned that the information plotted in Figure~\ref{xia}
comes from a Fisher matrix computed in the FKP approximation,
which is correct only for scales small compared to the scale of the survey.
The FKP approximation tends to misestimate, probably underestimate,
the information
contributed by regions near (within a wavelength of) survey boundaries,
since it assumes that those regions are accompanied by more correlated
neighbours than is actually the case.
The problem is most severe in surveys like LCRS,
where everyone lives near the coast.
Thus the information plotted in Figure~\ref{xia}
probably underestimates the true information
in the LCRS, especially at larger scales.

For a particular choice of resolution in wavenumber,
the information per unit log wavenumber translates into an
uncertainty in the corresponding uncorrelated band-powers.
The amount of information $\Delta I_k$ in a scaled band-power $\hat\bB_k/\bX_k$
of width $\Delta \ln k$ is equal to the corresponding diagonal element
$\bLambda_k$ of the Fisher matrix of the scaled band-powers:
\be
  \Delta I_k = {\dd I \over \dd\ln k} \Delta\ln k = \bLambda_k
  \ .
\ee
The inverse square root of this is the expected error in the
scaled band-power
\be
  \langle (\Delta \hat\bB_k/\bX_k)^2 \rangle^{1/2}
  = (\Delta I_k)^{-1/2}
  = \bLambda_k^{-1/2}
  \ .
\ee

Figure~\ref{xik}
illustrates an example of the error bars expected on the
decorrelated prewhitened nonlinear power spectrum of SDSS,
for the $\Lambda$CDM prior power spectrum.
As always in likelihood analysis,
`the errors are attached to the model, not to the data'.
Figure~\ref{xik} demonstrates that,
if baryonic wiggles are present at the expected level,
and if nonlinear evolution leaves at least the first wiggle intact,
as suggested by $N$-body simulations
(Meiksin, White \& Peacock 1999),
then SDSS should be able to recover them.
According to the prognostications of
Eisenstein, Hu \& Tegmark (1999),
the detection of baryonic features in the galaxy power spectrum
should assist greatly in the business of inferring cosmological parameters
from a combination of CMB and large scale structure data.

\section{Conclusions}

Amongst the infinity of possible ways to resolve the galaxy power spectrum
into decorrelated band-powers,
the square root of the Fisher matrix, or a scaled version thereof,
offers a particularly good choice.
The resulting band-powers are narrow, approximately symmetric,
and well-behaved in the presence of noise.

By contrast, a principal component decomposition
yields band-powers that are broad and wiggly,
which renders them of little practical utility.
A Cholesky decomposition of the Fisher matrix
works better than principal component decomposition,
but not as well as the square root of the Fisher matrix.
On the good side,
Cholesky band-power windows are narrow and not wiggly;
on the bad side,
Cholesky band-power windows are skewed to one side,
and they respond poorly to the presence of small negative eigenvalues in the
Fisher matrix,
as can occur because of a combination of numerical noise and
the various approximations that go into computing the Fisher matrix.

In summary, the square root of the Fisher matrix is a useful 
tool for decorrelating the power spectrum not only of
CMB fluctuations (Tegmark \& Hamilton 1998), but also of galaxy
redshift surveys.

We conclude with the caveat that this paper,
like the preceding Paper~3, has ignored redshift distortions,
and other complicating factors such as light-to-mass bias.
In any realistic analysis of real galaxy surveys,
such complications must be taken into account.

\section*{Acknowledgements}

We thank Huan Lin for providing FKP-weighted pair integrals for LCRS,
Michael Strauss for details of the angular masks
of SDSS and {\it IRAS\/} 1.2~Jy,
and David Weinberg for the radial selection function of SDSS.
This work was supported by
NASA grants NAG5-6034 and NAG5-7128
and by
Hubble Fellowship HF-01084.01-96A from STScI, operated by AURA, Inc. 
under NASA contract NAS5-26555.

%

\end{document}